\documentclass{ws-p9-75x6-50}

\begin{document}

\title{The Casimir force between metallic mirrors}

\author{Astrid Lambrecht, Cyriaque Genet and Serge Reynaud}

\address{Laboratoire Kastler Brossel
\footnote{mailto:lambrecht@spectro.jussieu.fr; 
http://www.spectro.jussieu.fr/Vacuum}\\Universit\'{e} 
Pierre et Marie Curie, Ecole Normale Sup\'{e}rieure et CNRS\\
Campus Jussieu, Case 74, 75252 Paris Cedex 05, France}

\maketitle

\abstracts{
In order to compare recent experimental results 
with theoretical predictions we study the 
influence of finite conductivity of metals on the Casimir effect. 
The correction to the Casimir force and energy due to imperfect reflection 
and finite temperature are 
evaluated for plane metallic plates where the dielectric functions of the
metals are modeled by a plasma model. The results are compared with the 
common approximation where conductivity and thermal corrections are evaluated 
separately and simply multiplied.}

\section{Introduction}
After its prediction in 1948 \cite{Casimir48} the Casimir force has been 
observed in a number of `historic' experiments \cite{Deriagin57,Sparnaay,Tabor68,Sabisky73}
but has only recently been remeasured with an improved experimental precision 
\cite{Lamoreaux97,Mohideen98,Roy99}. An accurate 
comparison with the predictions of Quantum Field Theory should therefore now
be possible, provided that 
theoretical predictions account for the differences between real
experiments and the idealized Casimir situation. In particular, experiments
are performed at room temperature between metallic mirrors while theoretical 
calculations are often performed at zero temperature and for perfect reflectors. 
As the experimental accuracy is claimed to be up to the order of 1\%,
theoretical expectations should also be computed with the same accuracy
if the aim is to test agreement between theory and experiment. A high accuracy 
is also important in order to control the effect of Casimir force when
studying small short range forces \cite{Fischbach,Carugno97,Bordag99}.

The influence of thermal field fluctuations on the Casimir force are known
to become important for distances of the order of a typical length 
\cite{Lifshitz,Mehra67,Brown69,Schwinger78} 
\begin{equation}
\lambda _{\rm T}=\frac{2\pi c}{\omega _{\rm T}}=\frac{\hbar c}{k_{B}T}
\label{lambdaT}
\end{equation}
When evaluated at room temperature, $\lambda _{\rm T}$ amounts to
approximately $7\mu $m. In contrast, the finite conductivity of metals has
an appreciable effect for distances smaller than or of the order of the
plasma wavelength $\lambda _{\rm P}$ determined by the plasma frequency $%
\omega _{\rm P}$ of the metal (see \cite{Lambrecht00} and references
therein) 
\begin{equation}
\lambda _{\rm P}=\frac{2\pi c}{\omega _{\rm P}}  \label{lambdaP}
\end{equation}
For metals used in the recent experiments, this wavelength lies in the range
0.1$\mu $m-0.2$\mu $m. This means that conductivity and thermal corrections
to the Casimir force are important in quite different distance ranges.

The purpose of this contribution is to give an accurate evaluation of the Casimir
force $F$ taking into account finite conductivity and temperature corrections at
the same time. To characterize the whole correction, we will compute the
factor $\eta _{\rm F}$ describing the combined effect of conductivity and
temperature 
\begin{eqnarray}
\eta _{\rm F} &=&\frac{F}{F_{\rm Cas}}  \nonumber \\
F_{\rm Cas} &=&\frac{\hbar cA\pi ^{2}}{240L^{4}}  
\label{Fcasimir}
\end{eqnarray}
$F_{\rm Cas}$ is the ideal Casimir force corresponding to perfect mirrors
in vacuum. $L$ is the distance between the mirrors, $A$ their surface and $%
\hbar $ and $c$ respectively the Planck constant and the speed of light.
We
will also evaluate the factors associated with each effect taken separately
from each other 
\begin{equation}
\eta _{\rm F}^{\rm P}=\frac{F^{\rm P}}{F_{\rm Cas}}\qquad \eta _{%
{\rm F}}^{\rm T}=\frac{F^{\rm T}}{F_{\rm Cas}}  \label{etaF}
\end{equation}
$F^{\rm P}$ is the Casimir force evaluated by accounting for finite
conductivity of the metals but assuming zero temperature and $F^{\rm T}$
is the Casimir force evaluated at temperature $T$ for perfect reflectors. 

Now the question is  
to which level of accuracy the complete correction factor $\eta _{\rm F}$ 
can be approximated as the product of the factors $\eta _{\rm F}^{\rm P}$ and 
$\eta _{\rm F}^{\rm T}$ ? To answer this question we will evaluate the quantity 
\begin{equation}
\delta _{\rm F}=\frac{\eta _{\rm F}}{\eta _{\rm F}^{\rm P}\eta _{%
{\rm F}}^{\rm T}}-1  \label{deltaF}
\end{equation}
which measures the degree of validity of the approximation where both
effects are evaluated independently from each other. 
We will give an analytical estimation of this deviation
in the end of this paper. We will 
also give the same results for the Casimir energy by defining a
factor $\eta _{\rm E}$ measuring the whole correction of Casimir energy
due to conductivity and temperature and then discussing the factors 
$\eta _{\rm E}^{\rm P}$ and $\eta _{\rm E}^{\rm T}$ and the deviation 
$\delta _{\rm E}$ in the same manner as for the force. 

In view of a comparison between experimental and theoretical results, it 
has to be noted that recent experiments
have not been performed in the plane-plane but in the plane-sphere configuration.
The Casimir force in this geometry is usually estimated from the proximity
theorem \cite{Deriagin68,Blocki77,Mostepanenko85,Bezerra97,Klimchitskaya99}.
Basically this amounts to evaluating the force by adding the contributions
of various distances as if they were independent. In the plane-sphere
geometry the force evaluated in this manner turns out to be given by the
Casimir energy evaluated in the plane-plane configuration for the distance $L
$ being defined as the distance of closest approach in the plane-sphere
geometry. Hence, the factor $\eta _{\rm E}$ evaluated here for the
energy can be used to infer the factor for the force measured in the
plane-sphere geometry. Surface roughness corrections will not be
considered in the following. Finally the dielectric response of the
metallic mirrors will be described by a plasma model. This model is known to
describe correctly the Casimir force in the long distance range which is
relevant for the study of temperature effects.

\section{Casimir force and free energy}

When real mirrors are characterized by frequency dependent reflection
coefficients, the Casimir force is obtained as an integral over frequencies
and wavevectors associated with vacuum and thermal fluctuations \cite
{Jaekel91}. The Casimir force is a sum of two parts corresponding to the 2
field polarizations with the two parts having the same form in terms of the
corresponding reflection coefficients 
\begin{eqnarray}
&&F=\sum_{k=-\infty }^{\infty } \frac {\omega _{\rm T}}{2}\ {\cal F}
\left[ k\omega _{\rm T}\right]   \nonumber \\
&&{\cal F}\left[ \omega \ge 0\right] =\frac{\hbar A}{2\pi ^{2}}\int_{%
\frac{\omega }{c}}^{+\infty }{\rm d}\kappa \ \kappa ^{2}\ f  \nonumber \\
&&f=\frac{r_{\bot }^{2}\left( i\omega ,i\kappa \right) }{e^{2\kappa
L}-r_{\bot }^{2}\left( i\omega ,i\kappa \right) }+\frac{r_{||}^{2}\left(
i\omega ,i\kappa \right) }{e^{2\kappa L}-r_{||}^{2}\left( i\omega ,i\kappa
\right) }  \nonumber \\
&&{\cal F}\left[ -\omega \right] ={\cal F}\left[ \omega \right] 
\label{Fexact}
\end{eqnarray}
$r_{\bot }$ (respectively $r_{||}$) denotes the amplitude reflection
coefficient for the orthogonal (respectively parallel) polarization of one
of the two mirrors. The mirrors are here supposed to be identical, otherwise 
$r_{\bot }^{2}$
should be replaced by the product of the two coefficients. $\omega $ is the
frequency and $\kappa $ the wavevector along the longitudinal direction of
the cavity formed by the $2$ mirrors. ${\cal F}\left[ \omega \right] $ is 
defined for positive frequencies and extended to negative ones by parity. 

The Casimir force (\ref{Fexact}) may also be rewritten after a Fourier 
transformation, as a consequence of Poisson formula \cite{Schwinger78}
\begin{eqnarray}
F &=&\sum_{m=-\infty }^{\infty } \widetilde{\cal F} 
\left( m\lambda _{\rm T} \right)   \nonumber \\
\widetilde{\cal F}(x) &=&\int_{0}^{\infty }{\rm d}\omega \ \cos
\left( \frac{\omega x}{c}\right) \ {\cal F}\left[ \omega \right] 
\label{Fpoisson}
\end{eqnarray}
The contribution of vacuum fluctuations, that is also the limit
of a null temperature $\left( \omega _{\rm T}\rightarrow 0\right) $ in
(\ref{Fexact}), corresponds to the contribution $m=0$ in (\ref{Fpoisson})
\begin{equation}
F^{\rm P}= \widetilde{\cal F} \left( 0 \right) =
\int_{0}^{\infty }{\rm d}\omega \ {\cal F}\left[ \omega
\right]   \label{Fplasma}
\end{equation}
Hence, the whole force (\ref{Fpoisson}) is the sum of this vacuum contribution 
$m=0$ and of thermal contributions $m \neq 0$.

We will consider metallic mirrors with the dielectric function $\varepsilon
\left( i\omega \right) $ for imaginary frequencies given by the plasma model 
\begin{equation}
\varepsilon \left( i\omega \right) =1+\frac{\omega _{\rm P}^{2}}{\omega^{2}}
\end{equation}
$\omega _{\rm P}$ is the plasma frequency related to the plasma wavelength 
$\lambda _{\rm P}$ by (\ref{lambdaP}). For the metals used in recent
experiments, the values chosen for the plasma wavelength $\lambda _{\rm P}$
will be 107nm for Al and 136nm for Cu and Au. 

We will also focus our attention on mirrors with a large optical thickness
for which the reflection coefficients $r_{\bot }\left( i\omega ,i\kappa
\right) $ and $r_{||}\left( i\omega ,i\kappa \right) $ correspond to a
simple vacuum-metal interface. Their expressions can be found in standard 
literature.

The Casimir energy will be obtained from the force by integration over the
mirrors relative distance 
\begin{equation}
E=\int_{L}^{\infty }F(x){\rm d}x
\end{equation}
As this procedure is performed at constant temperature, the energy thus
obtained corresponds to the thermodynamical definition of a free energy.
For simplicity we will often use the denomination of an energy. We will
define a factor $\eta _{\rm E}$ measuring the whole correction of energy
due to conductivity and temperature effects with respect to the ideal
Casimir energy 
\begin{eqnarray}
\eta _{\rm E} &=&\frac{E}{E_{\rm Cas}}  \nonumber \\
E_{\rm Cas} &=& \frac{\hbar c A \pi ^{2}}{720L^{3}}
\label{defEtaE}
\end{eqnarray}
The positive value of the energy here means that the Casimir energy is a binding
energy while the positive value of the force is associated with an
attractive character.

\section{Numerical evaluations}

In the following we present the numerical evaluation of the correction
factors of the Casimir force and energy using equations written in the
former section.

The force correction factor 
was evaluated for the experimentally relevant distance range of 
0.1-10$\mu$m with the help of equation (\ref{Fpoisson}), supposing explicitly 
a plasma model for the dielectric function, and the result was normalized by 
the ideal Casimir force. The energy correction factor was then calculated by 
numerically integrating 
the force and normalizing by the ideal Casimir energy.
Integration was restricted to a finite interval, the upper 
limit exceeding at least by a factor of $10^4$ the distance at which the 
energy value was calculated. Extending the integration range by a factor of 
100 changed the numerical result by less than $10^{-7}$.

\begin{figure}[tbh]
\centerline{\epsfig{figure=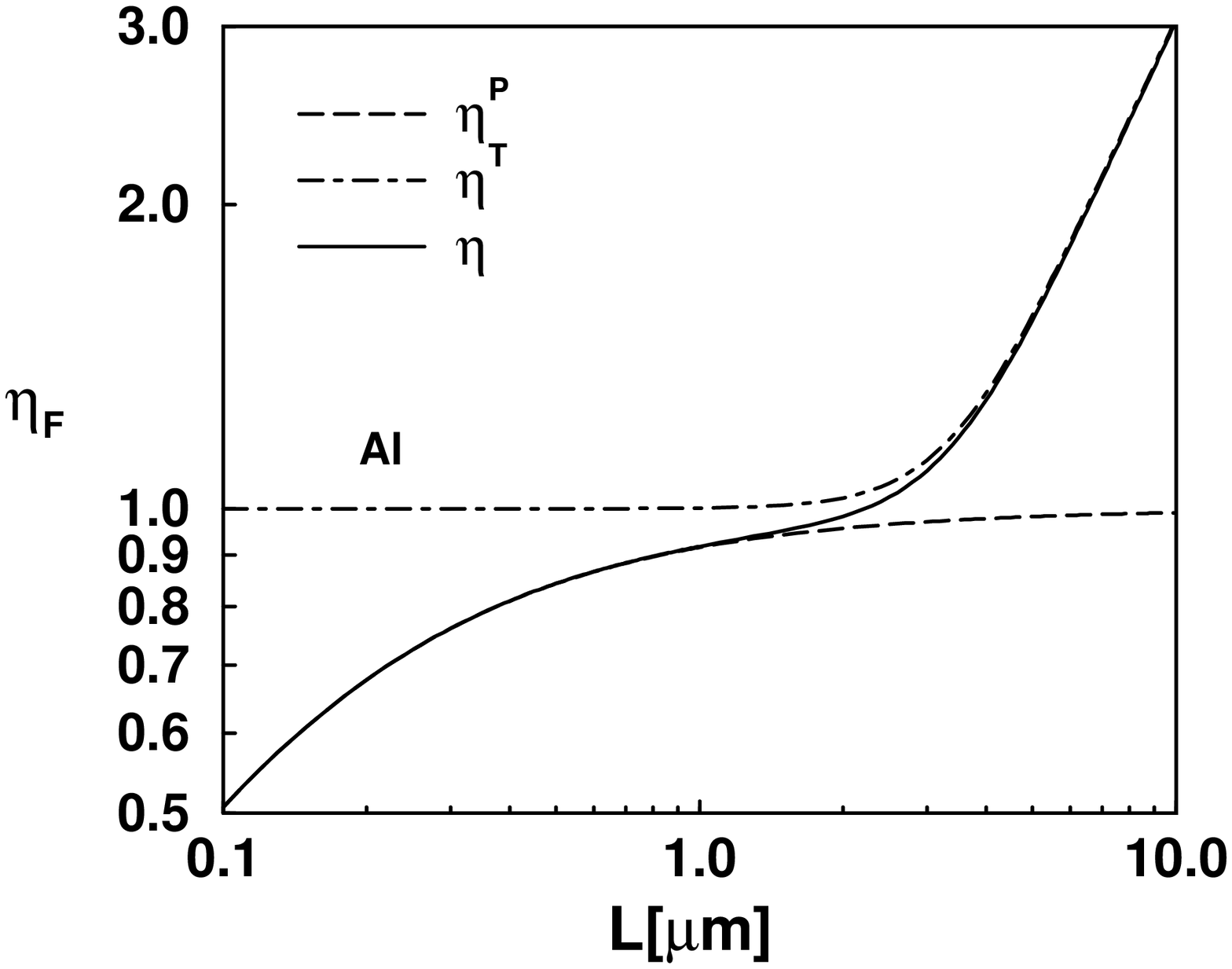,width=8cm}}
\centerline{\epsfig{figure=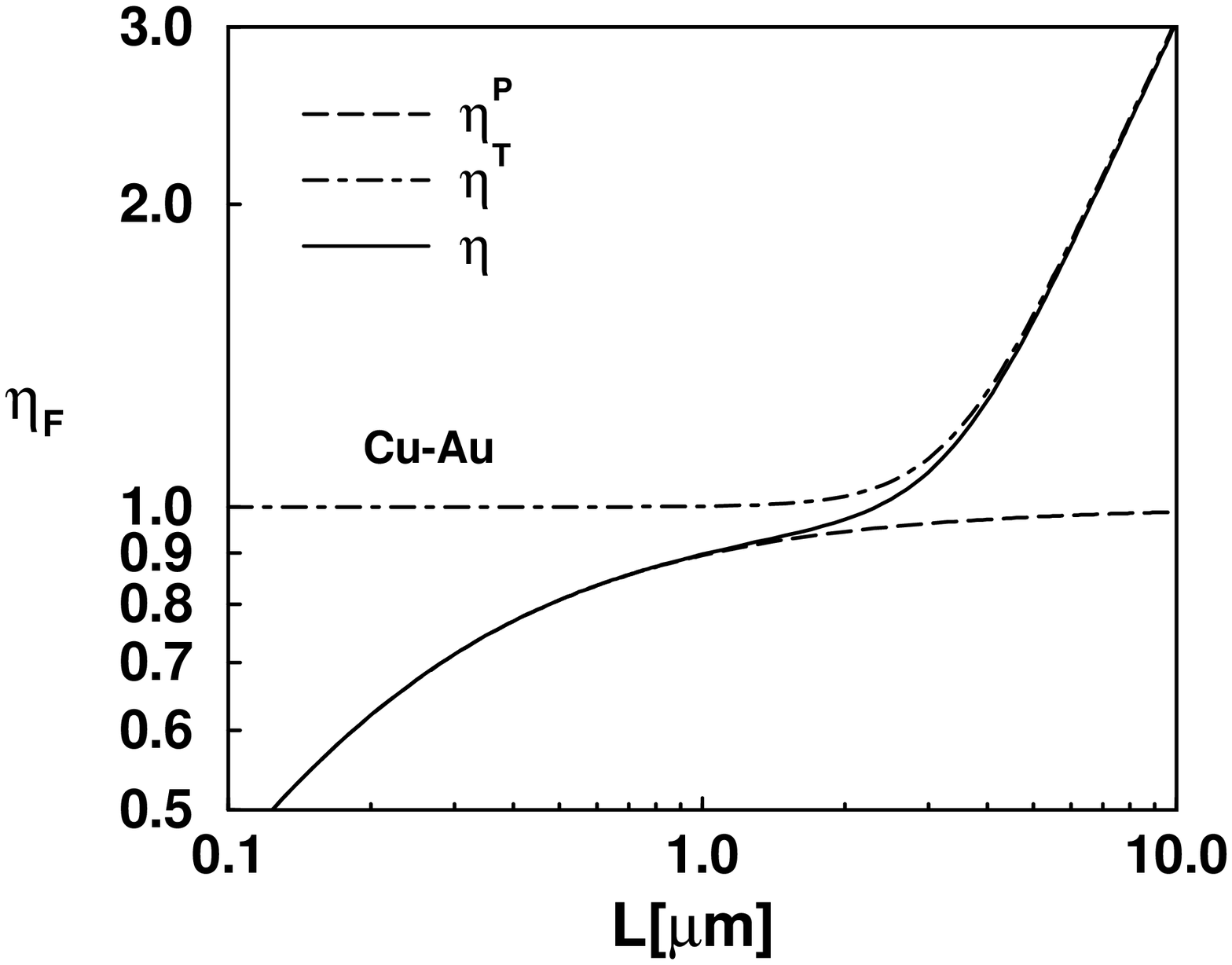,width=8cm}}
\caption{Force correction factor for Al (upper figure) and Cu and Au (lower
graph) as function of the mirrors distance at $T=300K$.}
\label{fig1}
\end{figure}

The results of the numerical evaluation of $\eta _{\rm F}$ are shown 
as the solid lines in figures \ref{fig1} for Al and for Cu-Au assuming a
temperature of $T=300K$. They are compared with the force reduction factor 
$\eta _{\rm F}^{\rm P}$ due to finite conductivity (dashed lines) and the
force enhancement factor $\eta _{\rm F}^{\rm T}$ calculated for perfect
mirrors at 300K (dashed-dotted lines).

\begin{figure}[tbh]
\centerline{\epsfig{figure=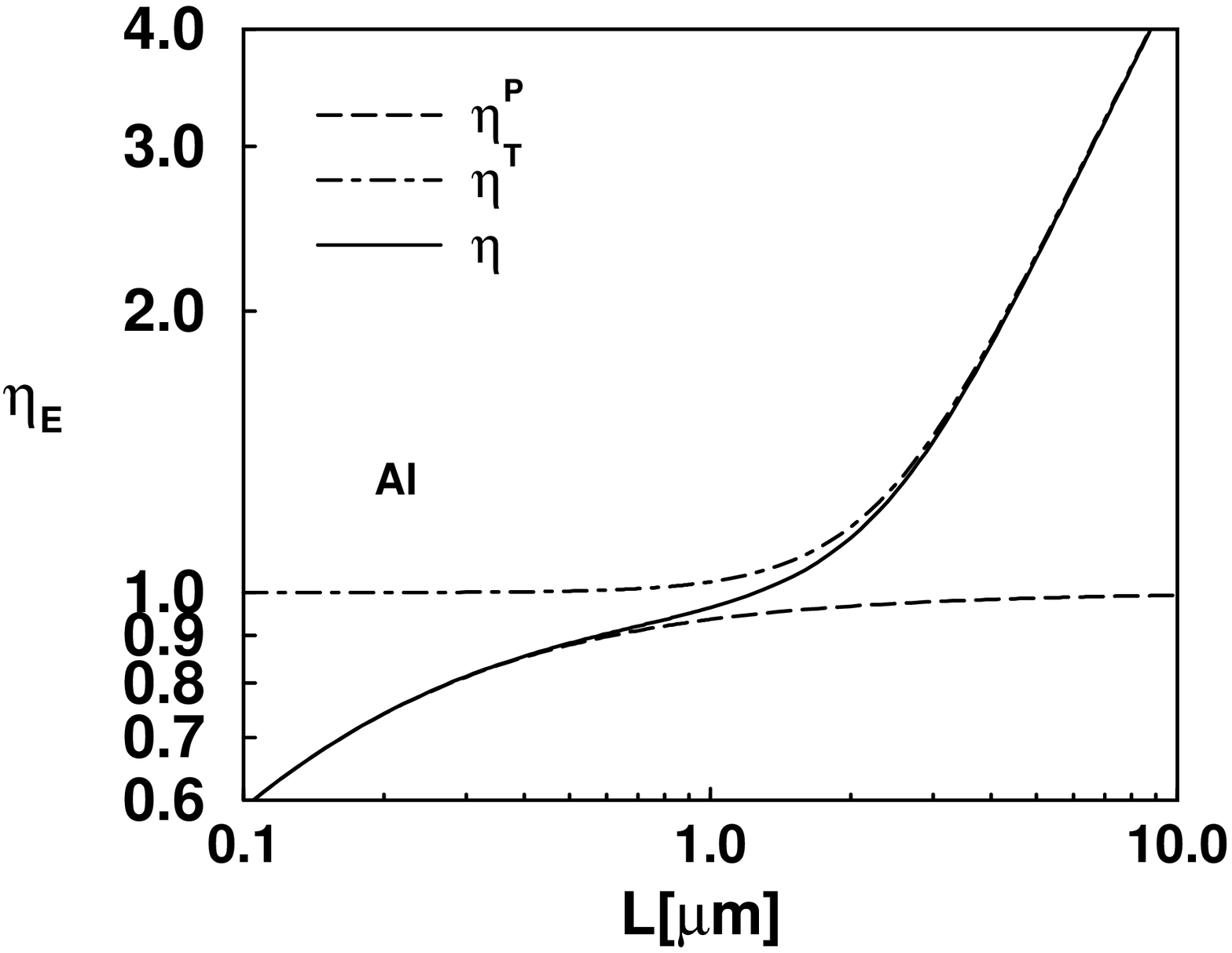,width=8cm}}
\centerline{\epsfig{figure=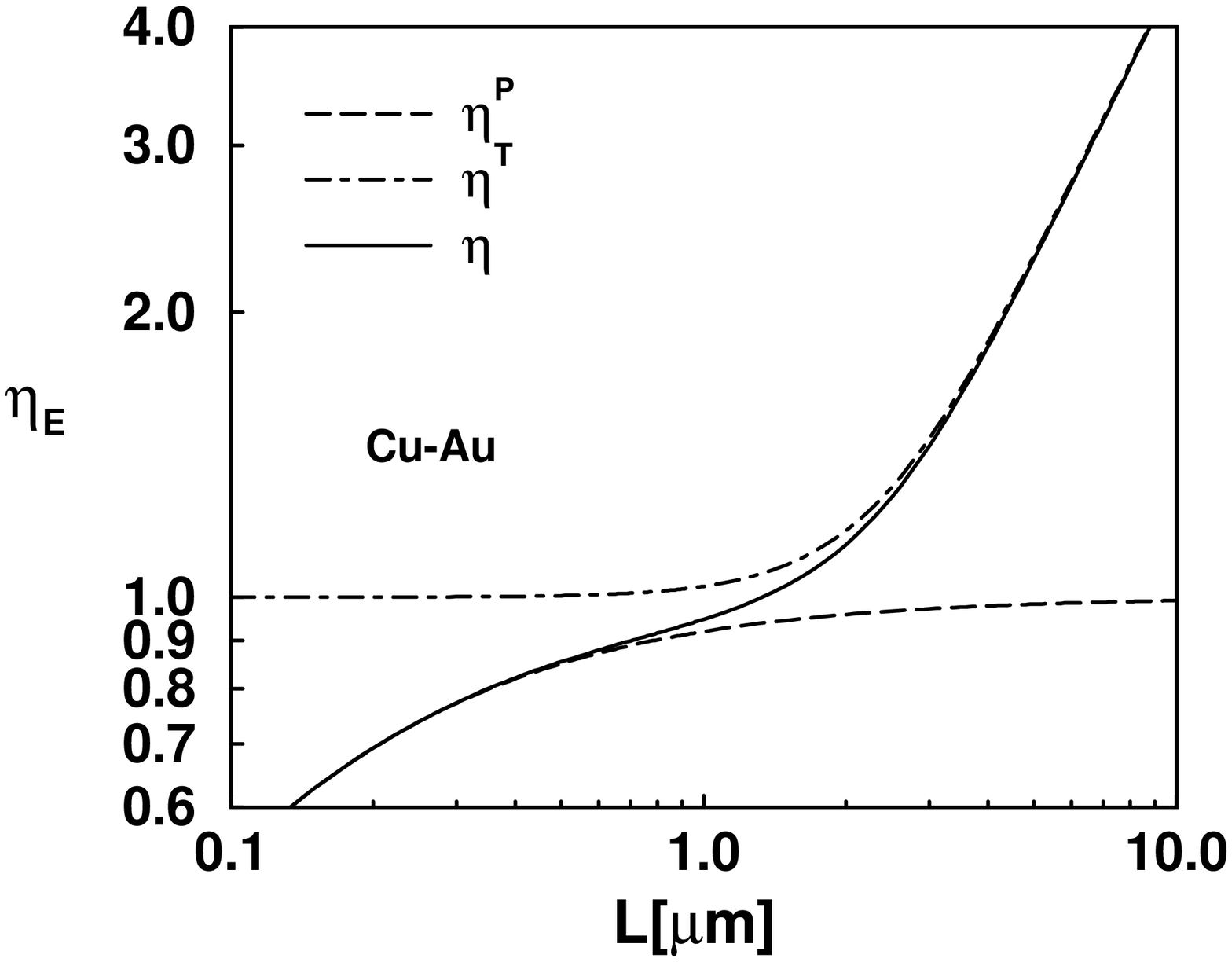,width=8cm}}
\caption{Energy correction factor for Al (upper figure) and Cu and Au (lower
graph) as function of the mirrors distance at $T=300K$.}
\label{fig2}
\end{figure}

Figure \ref{fig2} shows similar results for the factor $\eta _{\rm E}$
obtained through numerical evaluation of the Casimir free energy. The shape
of the graphs is similar to the ones of the force. However, while finite
conductivity corrections are more important for the force, thermal effects
have a larger influence on energy. 

\begin{figure}[tbh]
\centerline{\epsfig{figure=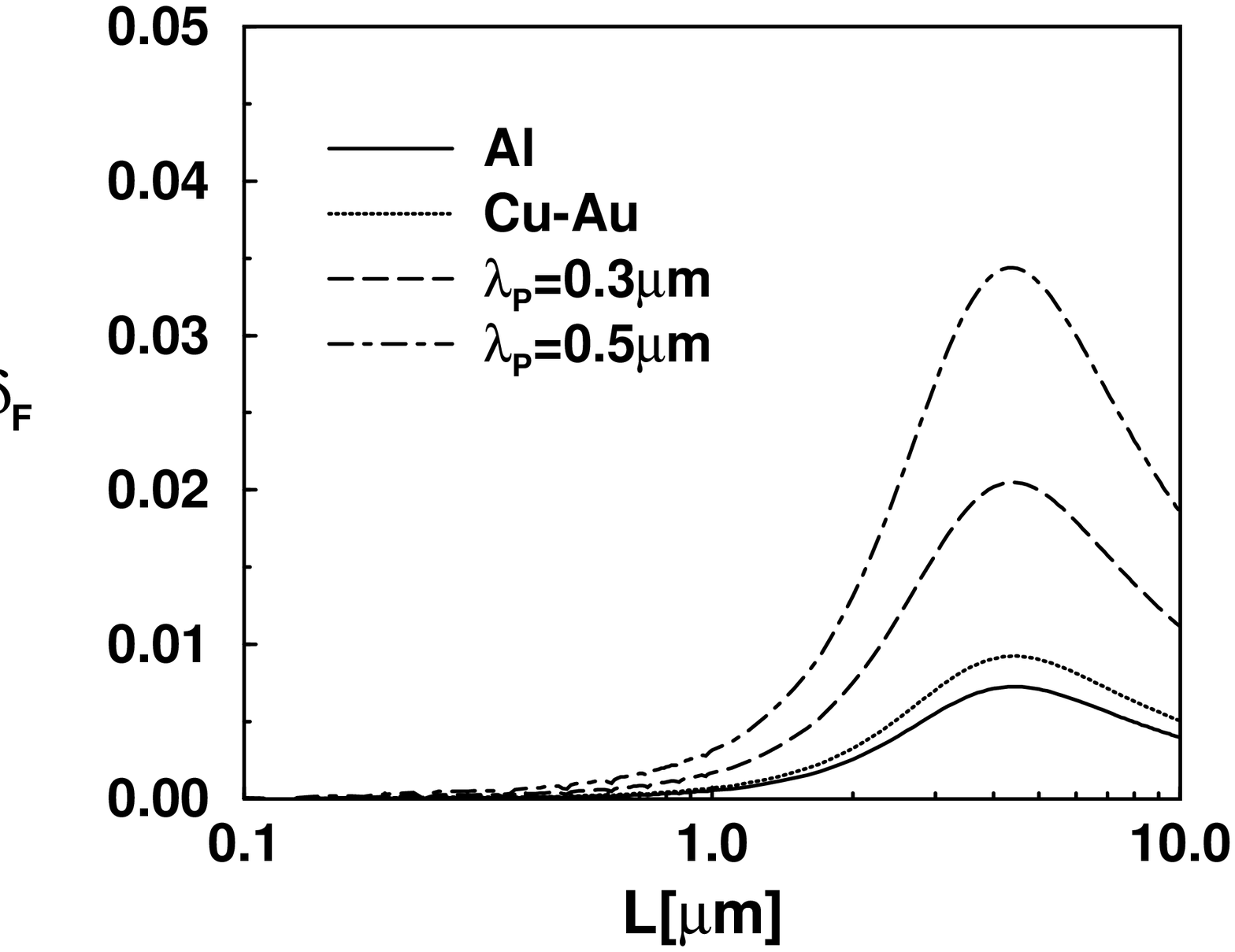,width=8cm}}
\centerline{\epsfig{figure=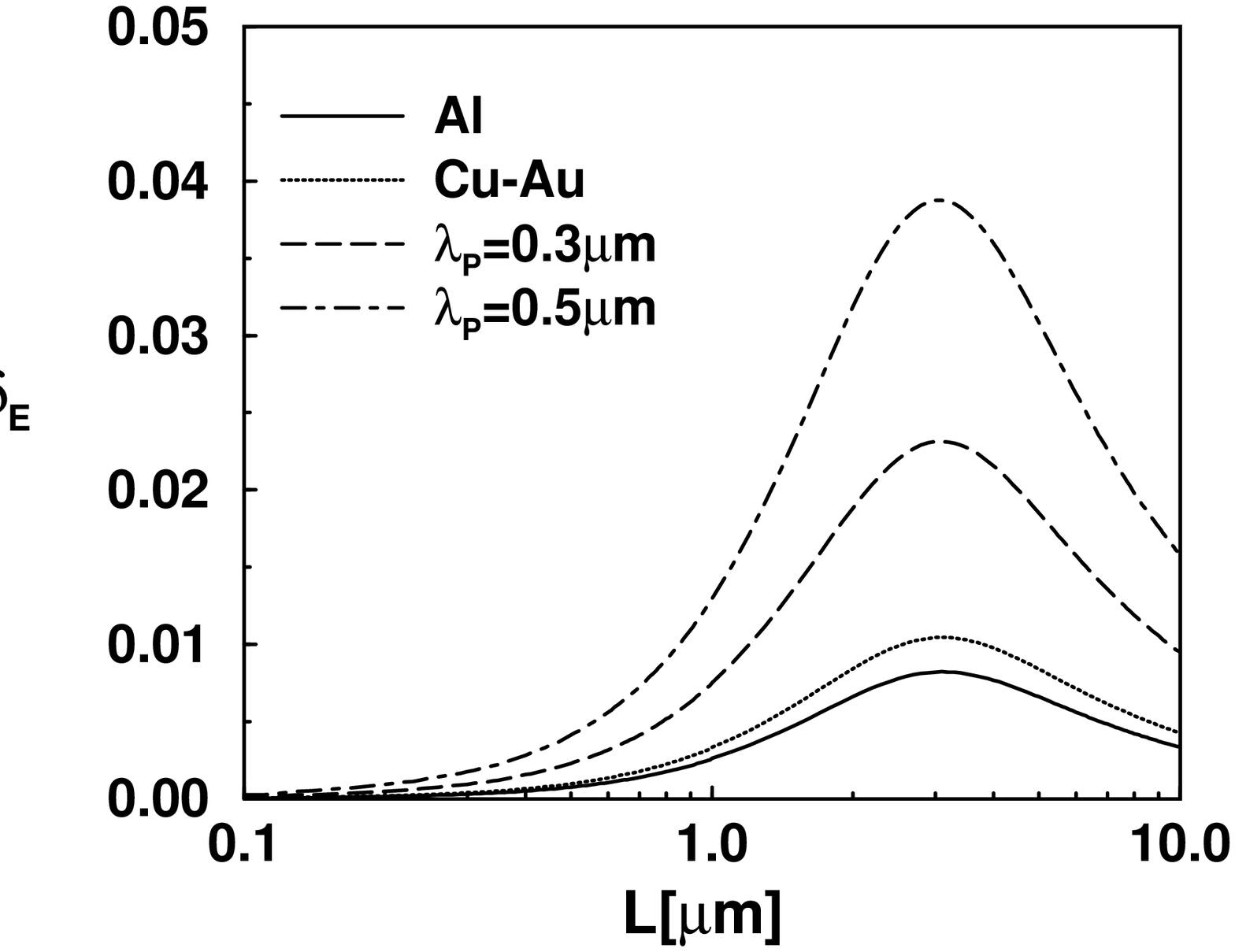,width=8cm}}
\caption{$\delta _{\rm F}$ (upper graph) and $\delta _{\rm E}$
(lower graph) as a function of the mirrors distance. The results are given
for the three metals Al, Cu-Au and two larger plasma wavelengths.}
\label{fig3}
\end{figure}

For the force as well as for the energy, temperature corrections are
negligible in the short distance limit while conductivity corrections may be
ignored at large distances. The whole correction factor $\eta $ behaves
roughly as the product $\eta ^{\rm P}\eta ^{\rm T}$ of the 2 correction
factors evaluated separately. However, both correction factors are
appreciable in the distance range $1-4\mu $m in between the two
limiting cases. Since this range is important for the comparison between
experiments and theory, it is necessary to discuss more precisely
how good is the often used approximation which identifies $\eta $ to the
product $\eta ^{\rm P}\eta ^{\rm T}$. In order to assess the quality of
this approximation, we have plotted in figure \ref{fig3} the quantities $%
\delta _{\rm F}$ and $\delta _{\rm E}$ (cf. eq.(\ref{deltaF}) as a function of 
the distance for
Al, Cu-Au and two additional plasma wavelengths. 
A value of $\delta =0$ would signify that the approximation gives an
exact estimation of the whole correction. An important outcome of our
calculation is that the errors $\delta _{\rm F}$ and $\delta _{\rm E}$
are of the order of 1\% for Al and Cu-Au at a temperature of $300K$. 

For estimations at the 5\% level, the separate calculation of $\eta ^{\rm P}$ 
and $\eta ^{\rm T}$ and the evaluation of $\eta $ as the product $\eta
^{\rm P}\eta ^{\rm T}$ can therefore be used. However, if a 1\% level
or a better accuracy is aimed at, this approximation is not sufficient. It
should be noticed furthermore that the error increases when the temperature
or the plasma wavelength are increased. It becomes of the order of 4\% for a
plasma wavelength of 0.5 $\mu $m at 300K. The sign obtained for $\delta $
means that the approximation gives too small values of force and energy.  

\section{Scaling laws for the deviations}

An inspection of figure \ref{fig3} shows that the curves corresponding to
different plasma wavelengths $\lambda _{\rm P}$ have similar shapes with a
maximum which is practically attained for the same distance between the
mirrors. The amplitude of the deviations, which is larger for the energy than for 
the force, is found to vary linearly as a function of the plasma wavelength 
$\lambda _{\rm P}$. 

\begin{figure}[tbh]
\centerline{\epsfig{figure=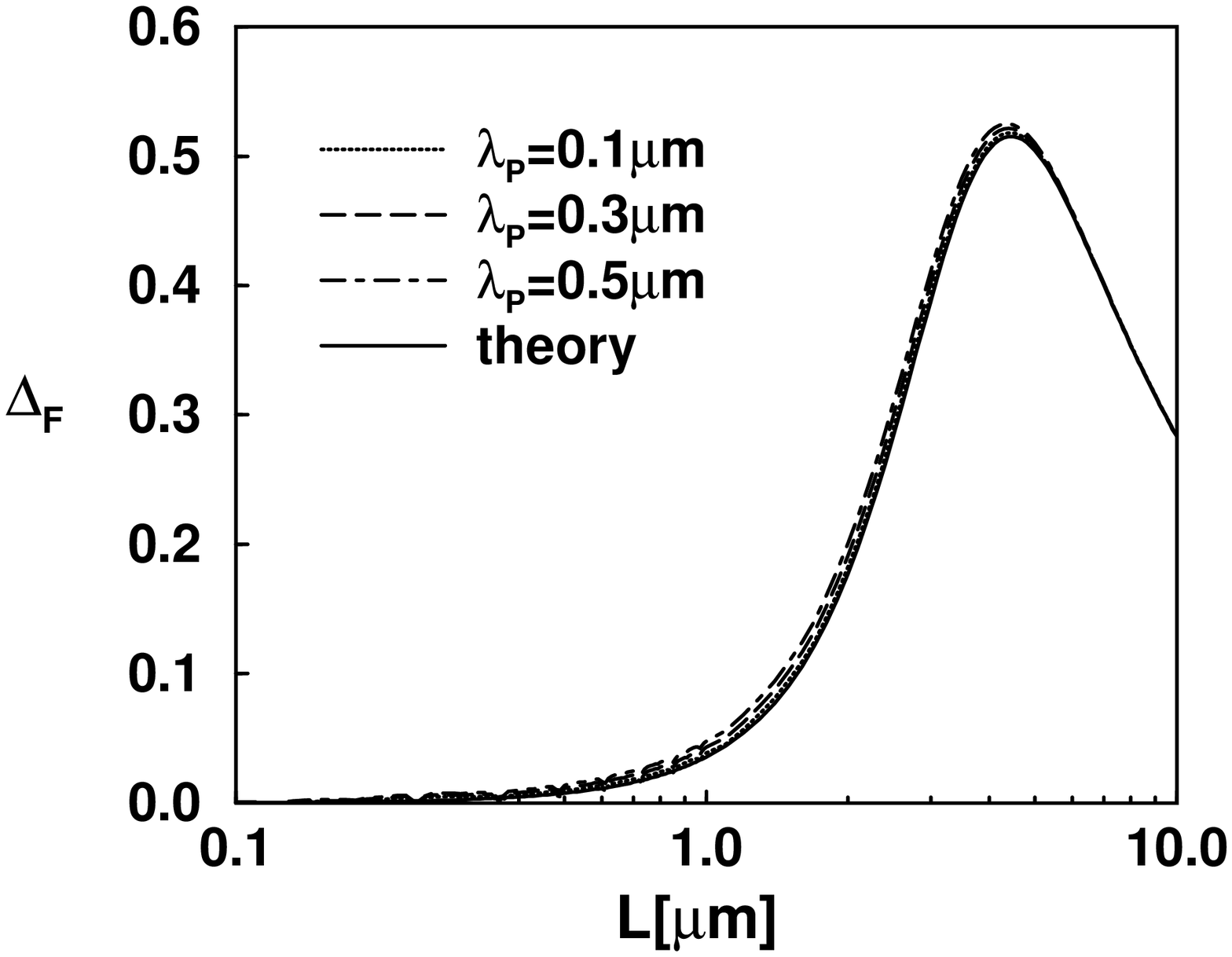,width=8cm}}
\centerline{\epsfig{figure=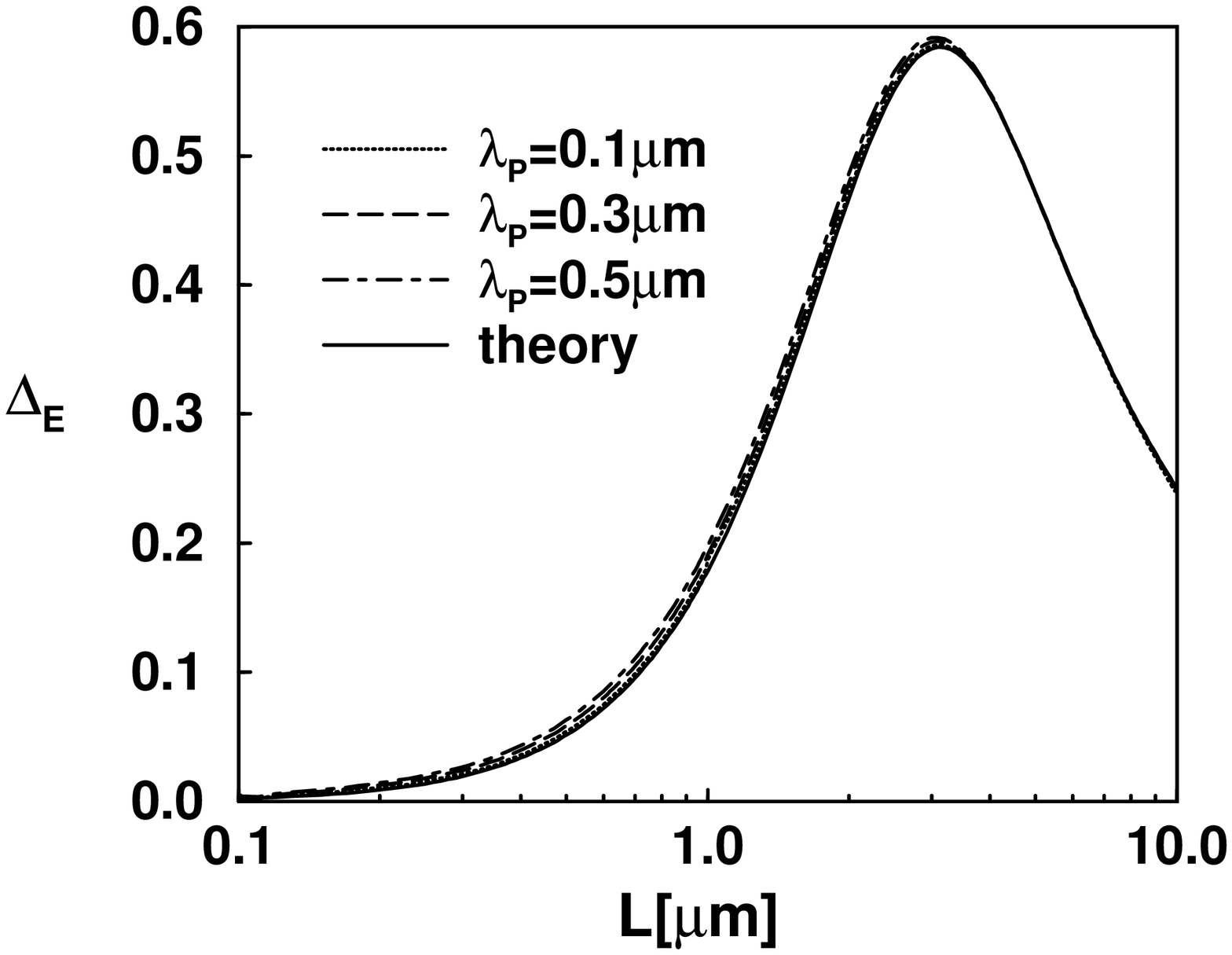,width=8cm}}
\caption{The deviations are represented for the force (upper
graph) and the free energy (lower graph) after the rescaling described
by equation (\ref{Scaling}). Different plasma wavelengths lead to nearly
identical functions, drawn as dotted, dashed and dotted-dashed lines. These
functions are hardly distinguishable from the solid lines which represent
the analytical expressions derived in the next section.}
\label{fig4}
\end{figure}

This scaling property is confirmed by figure \ref{fig4} where we have drawn
the deviations after an appropriate rescaling
\begin{equation}
\Delta =\frac{\lambda _{\rm T}}{\lambda _{\rm P}}\delta  \label{Scaling}
\end{equation}
The curves obtained for $\Delta _{\rm F}$ and $\Delta _{\rm E}$ for
different plasma wavelengths at temperature $T=300K$ are nearly perfectly
identical to each other. These curves correspond to values of the plasma 
wavelength small compared to the thermal wavelength and the scaling law
would not be obeyed so well otherwise. 

In other words, the deviations $\delta _{\rm F}$ and $\delta _{\rm E}$
are proportional to the factor $\frac{\lambda _{\rm P}}{\lambda _{\rm T}}
$ on one hand, and to the functions $\Delta _{\rm F}$ and $\Delta _{\rm E}$ 
on the other hand. The latter functions, which no longer depend on 
$\lambda _{\rm P}$, provide a simple method for reaching a good accuracy in
the theoretical estimation of the whole correction factor  
\begin{equation}
\eta =\eta ^{\rm P}\eta ^{\rm T}\left( 1+\frac{\lambda _{\rm P}}{%
\lambda _{\rm T}}\Delta \right) 
\end{equation}
This method is less direct than the complete numerical integration of the
forces which has been performed for obtaining the curves presented in the
previous section. But it requires easier computations while nevertheless 
giving accurate estimations of the correction factors. Typically, the 
deviation $\delta$ with a magnitude of the order of the \% may be estimated 
with a much better precision through the mere inspection of figure \ref{fig4}. 

One may explain this scaling law by
using a partial analytical integration of the whole correction factors and 
calculating analytical expressions for the functions $\Delta _{\rm F} $ and
$\Delta _{\rm E} $ to first order in $\lambda_{\rm P}$ :
\begin{eqnarray}
\Delta _{\rm F} &=& \frac{8}{3 \pi} \frac{\lambda _{T}}{L} 
\frac{\eta _{\rm F}^{\rm T}-1}{\eta _{\rm F}^{\rm T}}
+ \frac{\lambda _{T}}{L} \frac{\phi _{\rm F}}{\eta _{\rm F}^{\rm T}}\\
\Delta _{\rm E} &=& \frac{2}{\pi} \frac{\lambda _{T}}{L} 
\frac{\eta _{\rm E}^{\rm T}-1}{\eta _{\rm E}^{\rm T}}
+ \frac{\lambda _{T}}{L} \frac{\phi _{\rm E}}{\eta _{\rm E}^{\rm T}}
\end{eqnarray}
This function is plotted as the solid line on figure \ref{fig4} and it is
found to fit well the results of the complete numerical integration 
presented before. The detailed calculations may be found in \cite{LambrechtPRA00}.

\section{Summary}

We have given an accurate evaluation of the Casimir
force and Casimir free energy between $2$ plane metallic mirrors, taking
into account conductivity and temperature corrections at the same
time. The whole corrections with respect to the ideal Casimir formulas,
corresponding to perfect mirrors in vacuum, have been characterized by
factors $\eta _{\rm F}$ for the force and $\eta _{\rm E}$ for the
energy. These factors have been computed through a numerical evaluation of
the integral formulas. They have also been given a simplified form as a
product of $3$ terms, namely the reduction factor associated with
conductivity at null temperature, the increase factor associated with
temperature for perfect mirrors, and a further deviation factor measuring a kind of
interplay between the two effects. This last factor turns out to lie in the
1\% range for metals used in the recent experiments performed at ambient
temperature. Hence the conductivity and temperature corrections
may be treated independently from each other and simply multiplied for
theoretical estimations above this accuracy level. 

However, when accurate comparisons between experimental and theoretical
values of the Casimir force are aimed at, the deviation factor 
has to be taken into account in theoretical estimations. 
The deviation factor is appreciable for distances greater than the plasma
wavelength $\lambda _{\rm P}$ but smaller or of the order of the thermal
wavelength $\lambda _{\rm T}$. We have used this property to derive a
scaling law of the deviation factor. This law allows one to obtain a
simple but accurate estimation of the Casimir force and free energy through 
a mere inspection of figure \ref{fig4}. Alternatively one can use analytical
expressions which have been obtained through a first order expansion in 
$\lambda _{\rm P}$ of the thermal contributions to Casimir
forces and fit well the results of complete numerical integration. 

We have
represented the optical properties of metals by the plasma model. This model
does not lead to reliable estimations of the forces at small distances but
this deficiency may be corrected by using the real dielectric function of
the metals. This does not affect the discussion of the present paper, except
for the fact that the pure conductivity effect has to be computed through an
integration of optical data for distances smaller than 0.5$\mu $m. Finally 
surface roughness corrections, which have 
not been considered in the present paper, are expected to play a 
significant role in theory-experiment comparisons in the short distance range.

\end{document}